# Prediction of stent under-expansion in calcified coronary arteries using machine-learning on intravascular optical coherence tomography

Yazan Gharaibeh, PhD,[a,b] Juhwan Lee, PhD,[a] Vladislav N. Zimin, MD, PhD,[c] Chaitanya Kolluru, MS,[a] Luis A. P. Dallan, MD, PhD,[c] Gabriel T. R. Pereira, MD,[c] Armando Vergara-Martel, BS,[c] Justin N. Kim, BS,[a] Ammar Hoori, PhD,[a] Pengfei Dong, PhD,[d] Peshala T. Gamage, PhD,[d] Linxia Gu, PhD,[d] Hiram G. Bezerra, MD, PhD,[e] Sadeer Al-Kindi, MD, PhD,[c] David L. Wilson, PhD[a,f,*]

[a] *Department of Biomedical Engineering, Case Western Reserve University, Cleveland, OH, 44106, USA*
[b] *Department of Biomedical Engineering, The Hashemite University, Zarqa, 13133, Jordan*
[c] *Cardiovascular Imaging Core Laboratory, Harrington Heart and Vascular Institute, University Hospitals Cleveland Medical Center, Cleveland, OH, 44106, USA*
[d] *Department of Biomedical and Chemical Engineering and Sciences, Florida Institute of Technology, Melbourne, FL, 32901, US*
[e] *Interventional Cardiology Center, Heart and Vascular Institute, University of South Florida, Tampa, FL, 33606, USA*
[f] *Case Western Reserve University, Department of Radiology, Cleveland, OH, 44106, USA*

*Corresponding author: dlw@case.edu
Telephone number: 216-368-4099, fax: 216-368-4969

## ABSTRACT

**BACKGROUND** Careful evaluation of the risk of stent under-expansions before the intervention will aid treatment planning, including the application of a pre-stent plaque modification strategy.

**OBJECTIVES** It remains challenging to achieve a proper stent expansion in the presence of severely calcified coronary lesions. Building on our work in deep learning segmentation, we created an automated machine learning approach that uses lesion attributes to predict stent under-expansion from pre-stent images, suggesting the need for plaque modification.

**METHODS** Pre- and post-stent intravascular optical coherence tomography image data were obtained from 110 coronary lesions. Lumen and calcifications in pre-stent images were segmented using deep learning, and numerous features per lesion were extracted. We analyzed stent expansion along the lesion, enabling frame, segmental, and whole-lesion analyses. We trained regression models to predict the poststent lumen area and then to compute the stent expansion index (SEI). Stents with an SEI < or ≥ 80% were classified as "under-expanded" and "well-expanded," respectively.

**RESULTS** Best performance (root-mean-square-error = $0.04 \pm 0.02$ mm$^2$, $r = 0.94 \pm 0.04$, $p < 0.0001$) was achieved when we used features from both the lumen and calcification to train a Gaussian regression model for a segmental analysis over a segment length of 31 frames. Under-expansion classification results (AUC=$0.85 \pm 0.02$) were significantly improved over other approaches.

**CONCLUSIONS** We used calcifications and lumen features to identify lesions at risk of stent under-expansion. Results suggest that the use of pre-stent images can inform physicians of the need to apply plaque modification approaches.

# Introduction

Patients with inadequate stent expansion are at high risk for adverse outcomes, including stent thrombosis and in-stent restenosis [1]. Both are well-described complications that often cause acute coronary syndromes and, in the worst-case scenario, sudden cardiac death. Despite substantial improvements made in interventional procedures, stent design, drugs, and polymers as well as the adoption of therapeutic strategies, acute stent thrombosis and in-stent restenosis remain critical issues [2]. Once a stent is implanted in atherosclerotic tissue that is highly resistant to dilation, there is no opportunity to apply a treatment such as atherectomy and the only option is to rely on very high pressure balloon dilations. Careful evaluation of these risks of under-expansions before the intervention will aid treatment planning, including the potential application of a pre-stent plaque modification strategy. AI-enhanced software for predicting stent explanation could play a fundamental role in the clinical management of patients, with important implications regarding the choice of optimal medical therapy.

In percutaneous coronary intervention (PCI) planning, intravascular optical coherence tomography (IVOCT) is a useful tool for identifying lesion severity, reference vessel size, lesion length, and the extent of calcification [2], [3]. IVOCT-guided stent implantation has been shown to significantly improve clinical outcomes as compared with angiographic-only guidance. IVOCT imaging provides detailed evaluation of the morphology of the calcifications and stent deployment, such as stent expansions, malapposition, and stent edge dissection [4].

Moderate to severe calcification in the culprit vessel is a strong predictor of major adverse cardiovascular events after PCI [5], perhaps related to stent under-expansion due to severely calcified plaque and inadequate lesion preparation before stent implantation. Coronary calcification can impair device delivery and inhibit stent expansion [6]. In response to the need to predict under-expansion, Fujino et al proposed an IVOCT-based calcium score to predict stent expansion and the need for pre-stent plaque modification [4]. Matsuhiro et al investigated whether several calcium parameters were correlated with stent expansion in moderate calcified lesions assessed by IVOCT [7]. To predict the occurrence of stent under-expansion, Min et al developed preprocedural intravascular ultrasound–based models [3]. Unlike intravascular ultrasound, IVOCT can penetrate the calcifications to visualize their thickness [4], allowing for a more complete assessment of calcifications.

Our approach builds on our extensive experience in analyzing IVOCT images. Our group has applied machine learning [8], [9] and deep learning [10]–[21] methods on IVOCT images for segmentation and plaque characterization as well as for stent analysis [22]–[25]. Also, we have integrated computational fluid dynamics [26] with finite element method [27] to study the hemodynamic alternations following stenting.

In this report, we create a machine learning method to predict stent deployment from pre-stent IVOCT images of calcified plaques. Deep learning is used to segment calcifications and lumen, which are then used to feed a machine learning regression model to predict lumen areas. As the minimum stent expansion index (SEI) is used to classify under-expansions (SEI<0.8), we compute SEIs and report classification predictions. As far as we know, this is the first time that such an automated, comprehensive machine learning approach has been applied to this important problem in interventional cardiology. We compare our results to those from current state-of-the-art.

# Methods

## Study Population

This study was a retrospective single-center study conducted at the University Hospitals Cleveland Medical Center, Cleveland, Ohio, USA. A total of 805 patients with stable angina and documented ischemia or acute coronary syndrome who had undergone IVOCT-guided PCI were eligible for the study. Exclusion criteria were (1) ostial lesion, (2) inability to cross the lesions with the OCT catheter because of tortuosity and/or occluding thrombus, (3) bypass graft stenosis, (4) in-stent restenosis, or (5) chronic total occlusion. In addition, lesions without either pre-stent or final OCT, without any calcium by OCT, or treated with plaque modification methods (i.e., rotational, laser, or orbital atherectomy or laser angioplasty) were excluded from this study. Final analysis included 104 patients. This retrospective study was approved by the Institutional Review Board of University Hospitals Cleveland Medical Center (Cleveland, OH, USA).

## Image Acquisition and Stent Intervention

After the administration of 250 *mcg* of intracoronary nitroglycerine, coronary angiography was performed with 6–7 F catheters through radial or femoral access. PCI was performed according to standard techniques. The choice of stent lengths and diameter was at the discretion of the interventionalist performing the procedure. Only drug-eluting stents were used in this study. OCT imaging was conducted using the C7XR FD-OCT Imaging System (Abbott Vascular, Santa Clara, CA, USA) after an injection of nitroglycerin (100–200 g). OCT was performed with a Dragonfly OPTIS 2.7 F 135-cm. Blood clearance was achieved by non-diluted iodine contrast using ISOVUE-370 (iopamidol injection, 370 mg iodine/mL; Bracco Diagnostics Inc., Princeton, NJ, USA). Images were acquired with an automated pullback at a rate of 36 mm/s using survey mode (375 frames, 75 mm), frame rate of 180 frames/s, and axial resolution of 20 μm. Images were deidentified and submitted to the Core Laboratory for independent offline analysis. Analysts who were blinded to the patient and procedural information in the Core Laboratory analyzed the OCT data. The reference lumen area was recorded by OCT automated measures or calculated by tracing the luminal contour on the proximal and distal reference segments.

## Machine Learning Model Development

Our goal was to predict stent expansion, as assessed by minimal SEI, from pre-stent IVOCT images. Stents with minimum SEI (mSEI) < or ≥ 80% were classified as "under-expanded" and "well-expanded," respectively. We achieved this by first predicting the poststent lumen area from the baseline images. The overall flow of our approach is shown in Figure 1. We adopted three approaches to predict under-expansion: frame, segmental, and lesion. The Supplemental Appendix presents the details for each method. Patients' data (110 lesions from 104 patients) were divided into a training data set (78 lesions) and a held-out test set (32 lesions). Regression algorithms were developed using five-fold cross validation across training data, where the training data were divided into internal training and test sets.

Thirty-nine features were extracted from each lesion. From the elemental features, first-order statistics (minimum, maximum, mean, median, SD, skewness, and kurtosis) were obtained where applicable (Supplemental Appendix). A total of 238 features were obtained (168 two-dimensional features; 69 three-dimensional features). For feature reduction in the regression model, we used least absolute shrinkage and selection operator (LASSO) [28]. In addition, we manually selected intuitively important features and statistics (highlighted in the Supplemental Appendix, Supplemental Table 1) to train the regression models. We called this intuitively selected group the calcification lesion expansion (CLE) group. Various calcium phenotypes (calcific nodules, calcium protrusion, and superficial concentric sheet) were used as independent variables to examine the performance effect of the lesion level analysis.

## Effect of Calcification Phenotype on Lesion-Based Prediction

Our data included three types of calcifications based on the reports in the literature: [29], [30] calcified nodule (~13%), calcified protrusion (~23%), and superficial calcific sheet (~64%). Calcified nodule has an erupted volcanic shape that protrudes into the lumen (Supplementary Figure 5A). Calcified protrusion also protrudes into the lumen but without eruptive nodules ((Supplementary Figure 5B). Calcific sheet has no protrusion in the lumen (Supplementary Figure 5C). The calcification phenotype was assigned by visual examination of the frame with the mSEI in the lesion.

## Statistical Analysis

We used SPSS version 10.0 (SPSS, Chicago, IL, USA) to perform the statistical analyses to evaluate patient and lesion characteristics at baseline (Table 1). All values were expressed as mean ± standard deviation (continuous variables) or as counts and percentages (categorical variables). Continuous variables were compared using unpaired Student *t*-tests, and categorical variables were compared using chi-square statistics. A *p* value of <0.05 was considered to indicate statistical significance.

# Results

## Clinical Data and Processing

Table 1 summarizes the baseline characteristics and procedural findings of the study cohort. There were 104 patients with 110 lesions: 55 were deemed under-expanded and 55 were deemed well-expanded. Supplemental Figure 1 shows the registration of pre- and post-stenting IVOCT images. Supplemental Figure 2 shows the results of the deep learning segmentation of calcifications and extraction of calcification attributes (Supplemental Figure 3). There was very good agreement between the manual and automated assessments.

## Performance of Analysis Methods

The regression models trained using the CLE set had the best prediction results. Supplemental Figure 4 shows the mean AUCs of the four feature groups using linear regression (LR) and Gaussian process regression (GPR). Features selected based on LASSO improved the performance for both LR and GPR compared with training both models using all features. We experienced a slight degradation in performance when training both models using a subgroup of the best 20 features ranked by LASSO and based on CLE feature group. CLE features (features highlighted in Supplemental Table 1) provided the best performance for both models. Supplemental Table 2 shows the best 20 features of the CLE group ranked using LASSO.

Using our optimal method, we obtained a very good regression result (Figure 2). Figure 2A shows the excellent agreement against the actual measurements on the test data set (root-mean-square-error = 0.4±0.02 $mm^2$, $r$ = 0.94±0.04, $p < 0.0001$). Figure 2B shows the residuals plot for the comparison between measured and predicted SEIs. Residual analysis indicated a very small bias (−0.1±0.7 $mm^2$), and most of the measurements were included in the prediction interval. Figure 3 compares the actual minimum SEI (blue bars) and the predicted minimum SEI (orange bars) for each case in the test set. In the figure, the cyan and magenta boxes indicate the misclassification of five cases.

It is enlightening to examine the actual and predicted lumen areas after stenting. Figure 4 shows two cases: an under-expansion case with a heavily calcified lesion (left panel) and one with a well-expanded stent in a vessel with mild calcification (right panel). The 3D visualization (Figures 4A, E) and the longitudinal view (Figure 4B, F) show calcification distribution. Figures 4C and 4G show the vessel after stent implantation. Predicted and actual lumen area curves (Figures 4D and 4H) are remarkably close, and the predicted SEI is close to actual. Applying the threshold of 0.8, both the predication and actual measurement were classified as an under-expanded stent (left panel). The model predicted fairly closely the location of the frame with the minimum SEI. Right panel shows a well-expanded stent in the presence of little calcification. In this case, there was remarkable agreement between the predicted and actual lumen area curves (Figure 4H), and the SEIs were within 2%. Both the prediction and actual SEIs were consistent with a well-expanded stent. The effects of calcifications on stent expansion are depicted in Figure 5. The upper panel corresponds to the case in Figure 4A. Because of the presence of calcifications, the post-stent lumen areas were not enhanced after stenting (frames 50–70). The lower panel corresponds to Figure 4E with a well-expanded stent (frames 25–80).

Figure 6 summarizes the performance of our algorithm for predicting stent under-expansion in cases of held-out tests. With this classification, the differences were within 4% after converting the regression problem to the classification problem using an SEI of 0.8 as a cutoff value with a mean accuracy of 0.84 and mean AUC of 0.85, respectively, across the folds. This indicates that our model is suitably trained and reliable.

From the regression results, we obtained predicted SEI values and classified cases with stent under-expansion, corresponding to mSEI < 0.8. The values predicted by the frame-based approach showed good agreement against the actual measurements (root-mean-square-error = 0.12±0.01 $mm^2$, $r$ = 0.63±0.05, $p < 0.0001$, accuracy = 0.78±0.04, and AUC = 0.79±0.04). Other methods provided inferior results. For comparison, the other analysis methods showed that the lowest performance among all methods was that of the regression model at the lesion level (AUC = 0.73±0.02). This is because of the small number of cases with this approach. Supplemental Table 3 summarizes the performance metrics.

## Impact of Calcification Phenotype on Stent Expansion

We investigated the impact of calcification phenotype on stent expansion using our lesion-specific analysis. Supplementary Figure 7 shows the mSEI values for the three phenotypes of the calcified lesions. To predict a well-expanded stent, we retrained the lesion-specific model by adding calcification phenotype as an independent variable. As shown in Supplemental Table 3, the AUC improved from 0.73±0.02 to 0.76±0.02.

## Comparison With the Existing State-Of-The-Art Techniques

Our optimized segmental method provided significantly better prediction as compared with the seminal work by Fujino [4]. Figure 7 shows the receiver-operating characteristic curves for our method, Fujino's method, and Fujino's machine learning (ML), where we used features from their publication and applied machine learning on our data. Our method clearly provides superior results with a significant difference (p<0.005 when comparing our method against that of the previously described method).

## Discussion

Our segmental method very well predicts results after stenting from IVOCT images taken before stenting. Lumen areas are predicted remarkably well along the full length of a vessel (Figure 5), given an accurate SEI prediction. When we analyzed the misclassifications in detail (Figure 3), we found only one gross misclassification. SEIs are predicted well (Figure 4), and stent under expansion (SEI < 0.8) is predicted with an accuracy of $0.84 \pm 0.04$ (Supplemental Table 3).

The ability to accurately predict results after stenting from images taken before stenting is significant. Once a stent is deployed in the atherosclerotic tissue, it is very difficult or impossible to increase the expansion, even with a post-dilation balloon under high pressure. Hence, the accurate prediction of stent deployment from pre-stent imaging is very important when planning intervention treatment. Some of the options for lesion preparation include rotational and orbital atherectomy, cutting or scoring the balloon, acoustic shock wave, and/or balloon pre-dilation. Because special devices are costly and carry some potential risk, the precise prediction of their need is advisable.

To our knowledge, this is the first IVOCT-based machine learning assessment of stent under-expansion. Some findings follow. The segmental approach provided a much better performance than single-frame or lesion approaches did. This might be because the segmental approach inherently includes many more instances of regression learning as compared with the lesion-based approach. With many more training samples, the difference in performance might be reduced. The segmental approach performed better than the single-frame approach did, likely because a single-frame does not capture the effect of the extent of calcification. This rationale is evident in the optimal frame length consisting of 31 frames or 6.2 mm. We infer that smaller segmental lengths did not account for the full local biomechanics, and longer segments might have required more training samples. Although we tried multiple feature reduction techniques, we found that our manually selected features, the CLE feature group, consisting of both lumen and calcification attributes, provided the best predictions. Adding the calcification phenotype as a risk factor enhanced the prediction performance of the lesion-based method.

It is instructive to examine the most important features for predicting stent deployment. Supplemental Table 2 shows the top 20 CLE features ranked by LASSO for the prediction of stent under-expansion. The most important predictors are the calcification angle and area, as well as the lumen area and percentage area stenosis. Supplemental Figure 4 shows the impact of using features selected by LASSO to train regression models using the segmental analysis approach. First, we trained the LR model and GPR using the whole feature set. Then we applied LASSO to rank the best features and used them to train the same models. LASSO successfully improved the performance for both models. Training both models using the CLE features provided the best performance. After using features from the CLE group and selected by LASSO, the performance for both models were slightly degraded. The prediction of stent expansion is a complicated process that requires the inclusion of many parameters during the training phase.

We compared our study to the groundbreaking study of Fujino et al [4]. They created an IVOCT-based calcium scoring system that can be used to predict stent under-expansion in which scoring was based on a simplified analysis of the maximum angle, maximum thickness, and length of calcification. Our more complex approach provided significantly better prediction as compared with both their method and their features in a machine learning approach (Figure 7). They determined calcification attributes in a semiautomated manner, whereas our method is fully automated based on deep learning segmentation, providing results within seconds. It will be important to reproduce these results on larger data sets.

## Conclusion

Machine learning approaches can predict stent deployment. Because it is difficult or impossible to correct stent under-expansion after the stent is deployed, a method to predict stent deployment from pre-stent images is important when planning the intervention.

## Acknowledgments

This project was supported by the National Heart, Lung, and Blood Institute through grants NIH R01HL114406 and NIH R01HL143484. This work was also supported by American Heart Association grant 20POST35210974 to Juhwan Lee. This research was conducted in a space renovated using funds from an NIH construction grant (C06 RR12463) awarded to Case Western Reserve University. The content of this report is solely the responsibility of the authors and does not necessarily represent the official views of the National Institutes of Health. Grants were obtained via collaboration between Case Western Reserve University and University Hospitals of Cleveland. The veracity guarantor, Juhwan Lee, affirms to the best of his knowledge that all aspects of this paper are accurate.

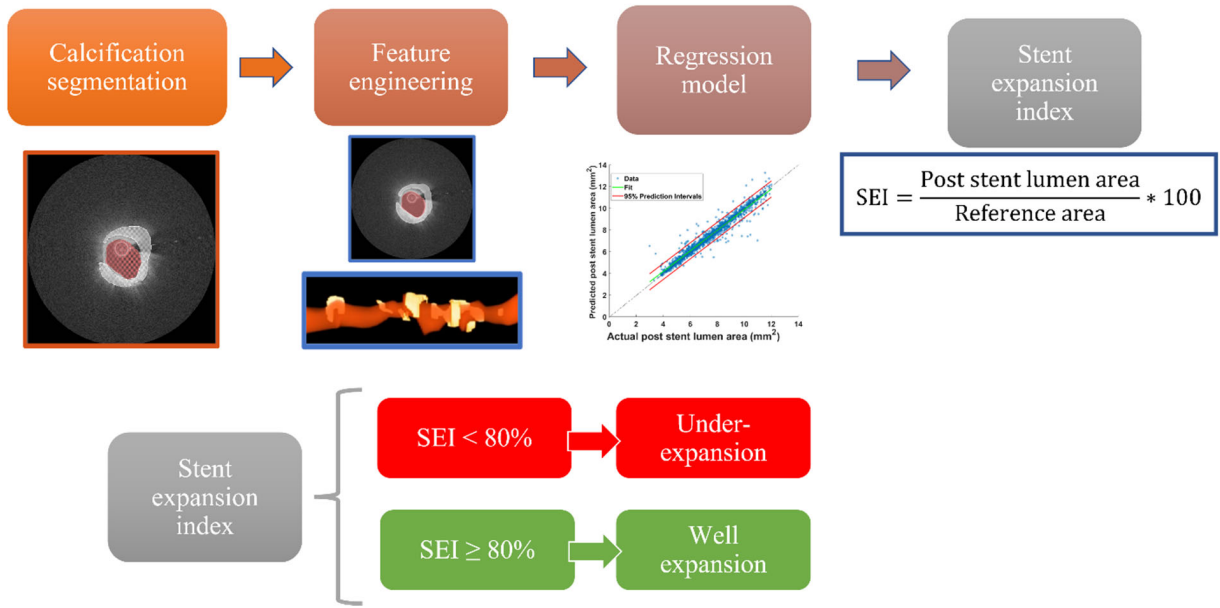

**Figure 1 Prediction of stent under-expansion workflow.** Features were extracted from the segmented image frames after automatic segmentation of both lumen and calcifications. Selected features were used to train regression models to predict the poststent lumen area along the vessel and compute the stent expansion index (SEI) for each stent. Stents with an SEI < or ≥80% were classified as "under-expanded" and "well-expanded," respectively.

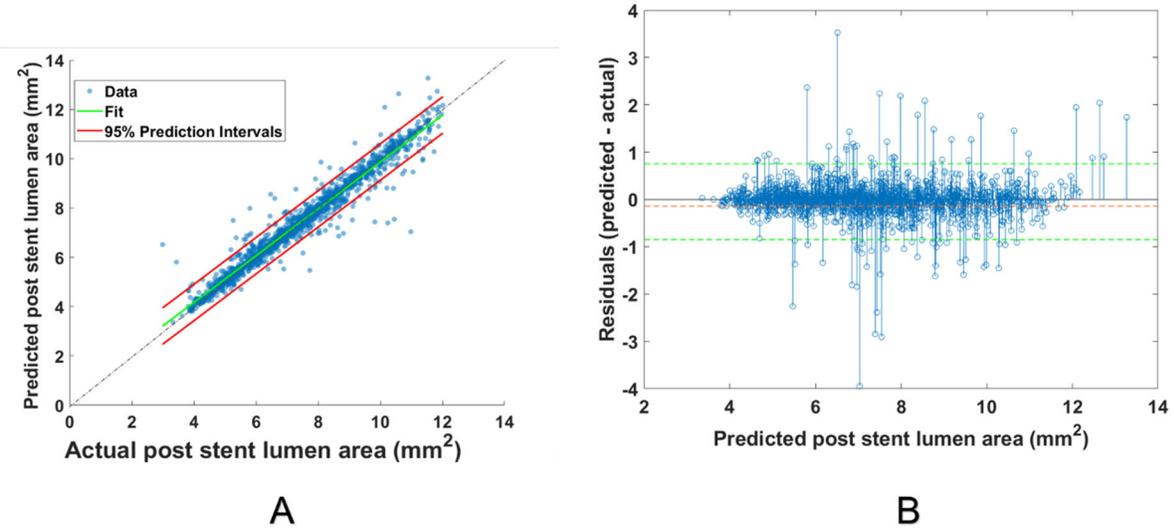

**Figure 2 Prediction of the lumen area for the segmental analysis method.** (A) The scatterplot shows a very high similarity ($r = 0.94\pm0.04$) between the predicted and measured area. (B) Residual plot, yielding a small bias and reasonable spread ($-0.1\pm0.7$ mm$^2$).

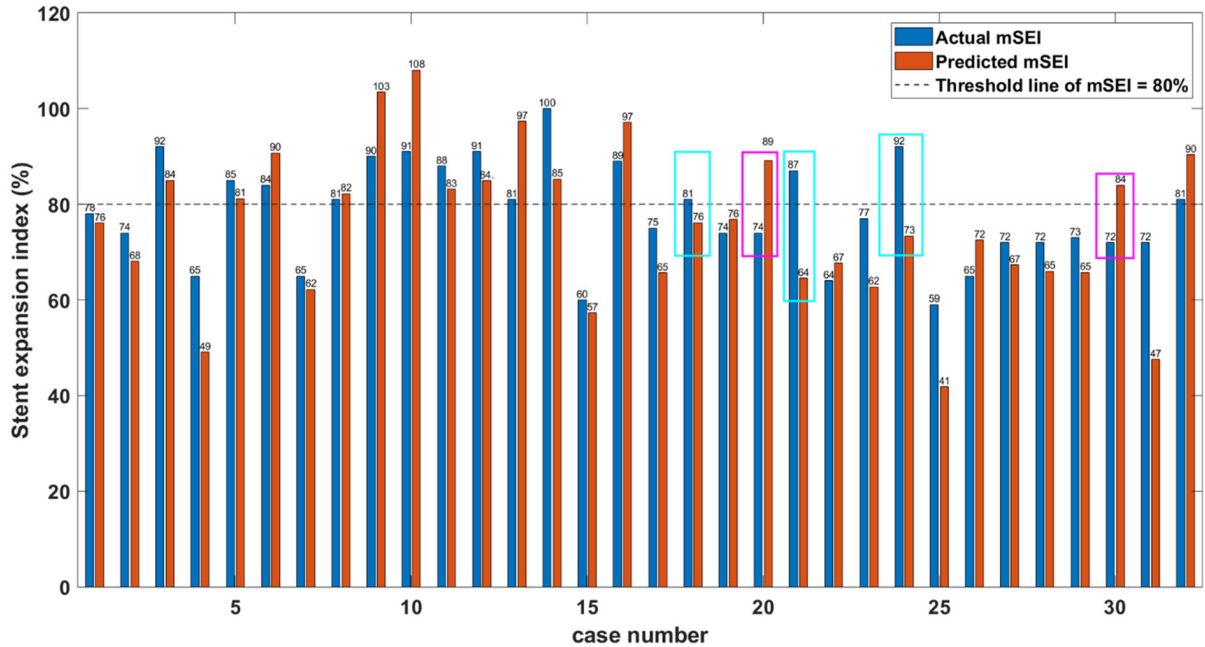

**Figure 3 Comparison between predict and actual expansion indices.** Predicted minimum SEI values (orange) are plotted with actual values (blue) for each stent in the test set. The horizontal line at 80% indicates the classification threshold for under-expansion. The cyan and magenta boxes indicate the five false-positive and false-negative cases, respectively. For two of the three false-positive cases, the prediction was sufficiently close to threshold, such that a physician might override the prediction after reviewing the case. For the two false-negative cases in which the software did not predict the need for plaque modification, the actual minimum SEI was 0.72 or better, not far from the acceptable threshold of 0.8. In some cases, there was a discrepancy between the mSEI values, as in case 25, but both were below the threshold of under-expansion. This difference is unimportant for clinical usage. Basically, there was only one gross error (case 21).

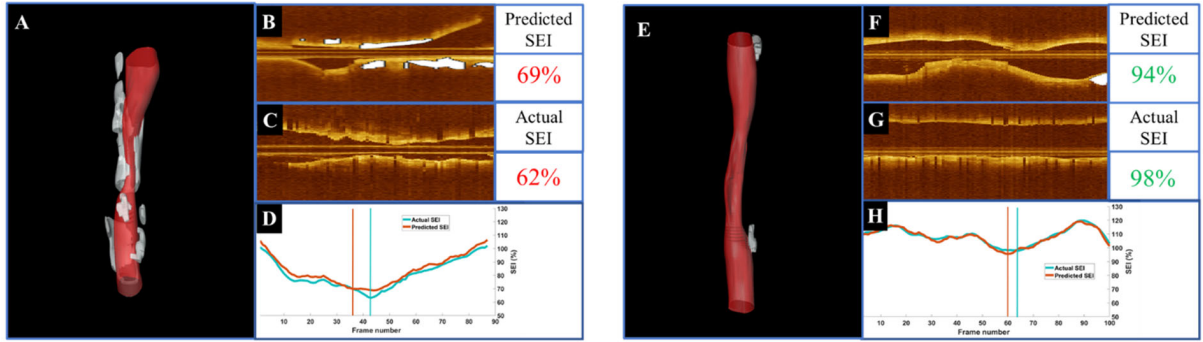

**Figure 4 Predicted stent area in cases with different calcifications severity.** Predicted stent area in a case of under-expansion in a heavily calcified lesion (left panel) and a case with a well-expanded stent in a vessel with relatively little calcification (right panel). (A, E) Three-dimensional rendering with calcifications in white, (B, F) longitudinal view before stenting with calcifications in white, (C, G) longitudinal view after stenting, and (D, H) predicted (orange) and actual (green) SEI following stenting. Our method predicted an SEI of 69%, which is close to the actual value of 62%, in which both values were indicative of under-expansion. The vertical bars in (D, H) show the locations corresponding to the minimum SEI values. The closeness of their location further suggests the predictive value of the regression model. The predicted and actual SEIs were 94% and 96%, respectively, and their locations were very close together.

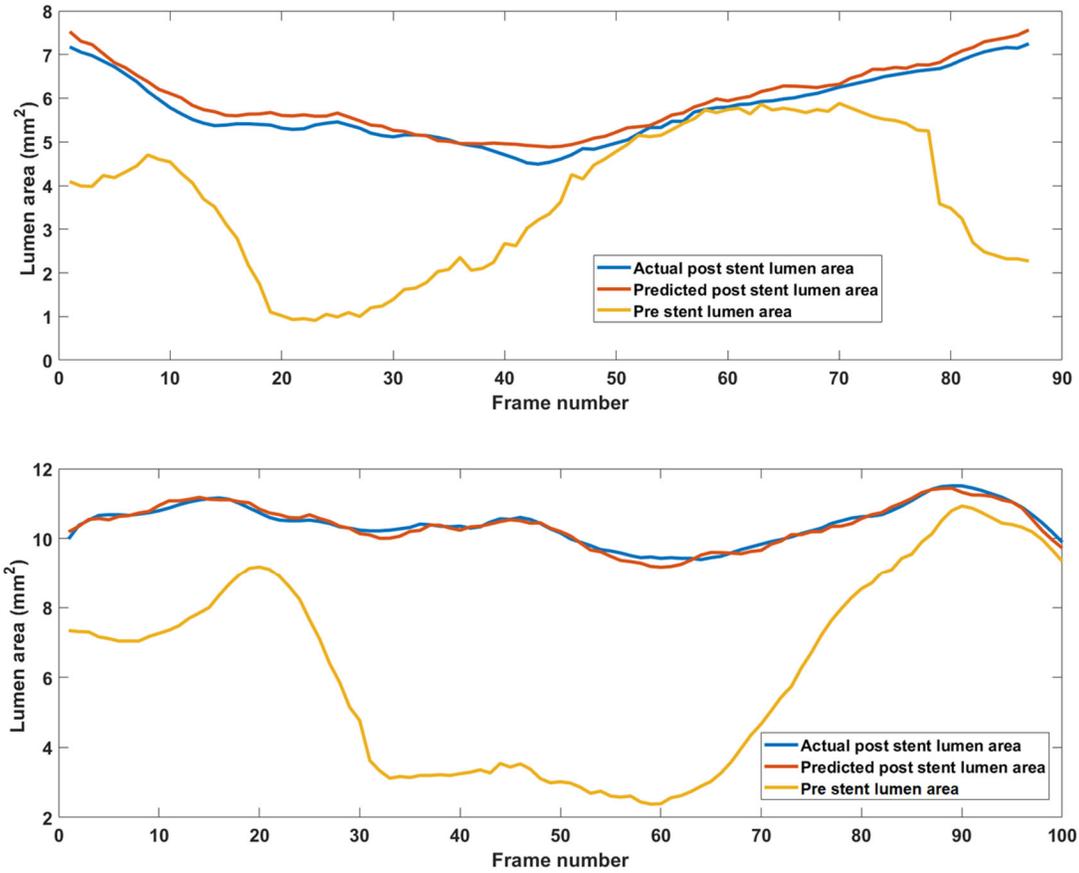

**Figure 5 Effect of calcifications on stent expansion.** Predicted and actual lumen areas after stenting are in blue and red, respectively. The orange curve represents the pre-stent lumen area for the registered pullback. The upper panel corresponds to the case in Figure 3A, which represents a case of under-expansion. Areas were not enhanced after stenting because of the presence of calcifications (frames 50–70). The lower panel corresponds to Figure 3E, with a well-expanded stent (frames 25–80).

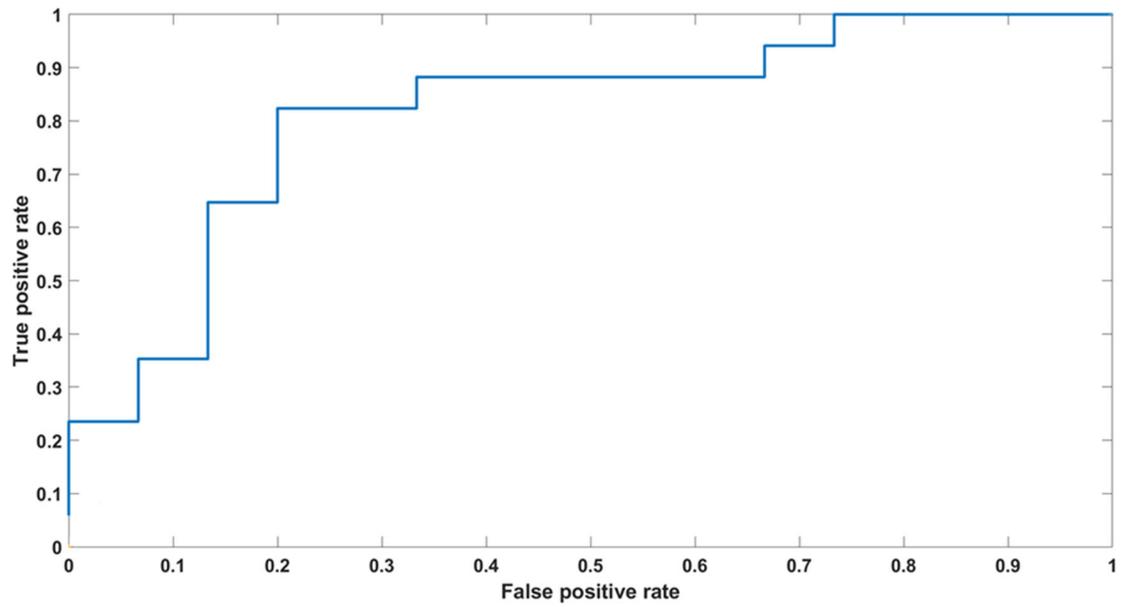

**Figure 6 Classification performance of the optimized segmental method.** Classification performance the receiver-operating characteristic curve (AUC of 0.84±0.02) after converting the regression results to the classification of stent under-expansion, using an SEI of 0.8 as the cutoff.

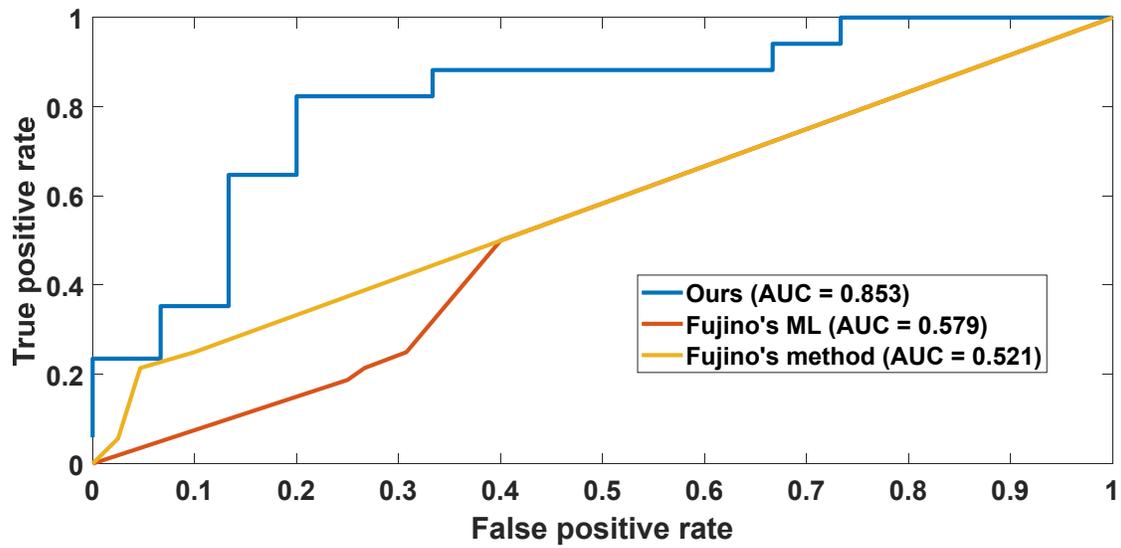

**Figure 7 Comparing the performance of our method to the state-of-the-art method.** Receiver-operating characteristic curves comparing the performance of our method to the state-of-the-art method reported by Fujino et al.*[4]* Shown are our method, Fujino's method as reported in *[4]*, and Fujino's ML. The AUCs were 0.853 (our method), 0.58 (Fujino's ML), and 0.52 (Fujino's method). Our method provided significantly improved prediction as compared with their method ($p<0.005$).

**Table 1 Baseline Patient Characteristics**

|  | Overall<br>n = 104 (100%) | Training /testing set<br>n=72/104 (69.2%) | Validation set<br>n=32/104 (30.8%) | p-value |
|---|---|---|---|---|
| **Patient-level characteristics** | Mean ± SD (N)<br>(Min, Max) | | | |
| Age (years) | 67.1±12.0<br>(104) [30, 98] | 67.6±12.4<br>(72) [40,98] | 65.9±10.9<br>(32 [30, 85]) | 0.494 |
| **Gender** | | | | |
| Male | 74/104 (71.15%) | 47/72 (65.3%) | 27/32 (84.4%) | |
| **Physical Measurement** | | | | |
| Height (cm) | 171.8 ± 9.8<br>(104) [144.8, 195] | 171.3±9.6<br>(72) [152.4, 195] | 172.8±10.1<br>(32) [144.8, 190.5] | 0.482 |
| Weight (kg) | 93.7 ± 25.3<br>(104) [40.8, 225] | 91.1±18.5<br>(72) [51.7, 151] | 99.0±35.9<br>(32) [40.8, 225] | 0.257 |
| BMI (kg/m$^2$) | 31.73 ± 8.1<br>(104) [17.9, 75.2] | 31.2±6.3<br>(72) [17.9, 49.1] | 32.9±11.3<br>(32) [19.4, 75.2] | 0.426 |
| **Medical History** | | | | |
| Hypertension | 99/104 (95.19%) | 69/72 (95.8%) | 30/32 (93.7%) | 0.676 |
| Diabetes Mellitus | 56/104 (53.84%) | 39/72 (54.2%) | 17/32 (53.1%) | 0.923 |
| Hyperlipidemia | 90/104 (86.54%) | 63/72 (87.5%) | 27/32 (84.4%) | 0.683 |
| Previous PCI | 8/104 (7.69%) | 6/72 (8.3%) | 2/32 (6.2%) | |
| Previous Myocardial Infarction | 60/104 (57.69%) | 42/72 (58.3%) | 18/32 (56.5%) | 0.062 |
| Heart Failure, LVEF <30% | 58/104 (55.77%) | 41/72 (56.9%) | 17/32 (53.1%) | 0.723 |
| Previous CABG | 8/104 (7.69%) | 6/72 (8.3%) | 2/32 (6.2%) | 0.703 |
| Current Smoker (≤6 Months) | 53/104 (50.96%) | 40/72 (55.5%) | 13/32 (40.6%) | 0.165 |
| Renal Dysfunction (Serum Creatinine > 2.0) | 53/104 (50.96%) | 34/72 (47.2%) | 19/32 (59.4%) | |
| Hemodialysis or Renal Transplant | 12/104 (11.54%) | 9/72 (12.5%) | 3/32 (9.4%) | |
| **Pre-procedure Presentation** | | | | |
| STEMI/Cardiogenic shock | 10/104 (9.6%) | 6/72 (8.3%) | 4/32 (12.5%) | |
| NSTEMI/Unstable Angina | 35/104 (33.65%) | 24/72 (33.3%) | 11/32 (34.4%) | |
| Stable Angina | 57/104 (54.81%) | 41/72 (56.9%) | 16/32 (50.0%) | |
| Silent Ischemia | 0/104 (0.0%) | 0/72 (0.0%) | 0/32 (0.0%) | |
| Other* | 1/104 (0.96%) | 1/72 (1.4%) | 0/32 (0.0%) | |

*Aortic stenosis.

# Supplemental Appendix

## Supplemental methods

### Machine learning model development

Our goal is to predict stent deployment, as assessed by the stent expansion index (SEI), from the present IVOCT images. The SEI for each frame is

$$\text{SEI(f)} = \frac{\text{post stent lumen area (f) (mm}^2)}{\text{mean of proximal and distal references(mm}^2)} * 100$$

The minimum SEI value (mSEI) is often reported as a metric of stent expansion. Consistent with the literature [31], we deem a stent "under expanded" if the mSEI is ≤80% and "well expanded" if the mSEI is >80%. Proximal and distal references were measured at the site with the largest lumen within 5 mm proximal and distal to the stented segment.

We adopted three approaches to predict under expansion: frame, segmental, and lesion. In the frame-based approach, we extracted features from each single-labeled frame to train a regression model to predict the poststent lumen area for each frame in a lesion. Features were from two-dimensional lumen and calcification feature groups. In the segmental approach, we extracted features from a moving segment of image frames across the lesion. We applied moving segments with different lengths (i.e., 3, 7, 15, 31, and 63 frames) and a stride of 1 frame. The poststent lumen area was predicted for the central frame. In the lesion-based approach, all features were computed from the target lesion. Stents with an SEI ≤ or >80% were classified as "under expanded" and "well expanded," respectively.

Patient data (110 lesions from 110 patients) were divided into a training data set (78 lesions) and held-out data set (32 lesions). Regression algorithms were developed using fivefold cross validation across training data, where the training data were divided into internal training and test sets. Before testing on held-out data, we trained the regression models across all training data. Image processing and network training were performed using MATLAB software package (R2021a, MathWorks Inc.) on a NVIDIA GeForce TITAN RTX GPU with 120 GB of RAM installed in a Dell Precision T7610.

### IVOCT feature extraction and selection

Features were extracted using in-house developed software. Thirty-nine features, from four feature groups (12 two-dimensional lumen features, 6 three-dimensional lumen features, 12 two-dimensional calcification features, and 9 three-dimensional calcification features) were extracted. First-order aggregation statistics (minimum, maximum, mean, median, SD, skewness, and kurtosis) were obtained where applicable (e.g., for two-dimensional features in the case of the segmental approach). A total of 238 features were obtained (i.e., 168 two-dimensional features; 69 three-dimensional features). Feature values were normalized between 0 and 1. Other features such as lumen area were not normalized, because the absolute area is actually important. Supplementary Table 1 summarizes the list of features. As appropriate, subsets of features were used for each of the frame, segmental, and lesion analysis approaches.

We used the least absolute shrinkage and selection operator (LASSO) [28] for feature reduction in the regression models. This selection method applied a shrinking (regularization) process in which it assigned weights to regression variables. LASSO shrinks the regression coefficients toward 0 to eliminate irrelevant features from the regression model. In addition, we manually selected intuitively important features and statistics (highlighted in Supplementary Table 1) to train the regression models. We called this intuitively selected group the calcification lesion expansion (CLE) group. We used LASSO to rank the CLE features based on their effects on the prediction. Calcification type (calcified nodule, calcified protrusion, and superficial calcific sheet) was used as an independent variable to examine the effect on the performance of the lesion-based analysis.

These groups of features were then evaluated by using each of the machine learning regression models (decision tree, regression support vector machine, Gaussian process regression, and ensemble models). The poststent lumen area was predicted using these models, and then the associated SEI was computed. The root mean square error was the performance metric for the regression prediction. We used lumen area regression predictions to compute SEI(f) for a given lesion. We then searched the SEI(f) values to obtain the mSEI. Predictions were deemed under expanded or well expanded based on the definitions above. Classification performance was assessed using ROCs and confusion matrices with associated statistics.

### Registration and Segmentation

The cross-sectional IVOCT frames between pre- and poststent pullbacks were co-registered frame by frame by using the landmarks such as the ostium, the opening of the side branch, coronary vein, calcifications, and other representative structures, as shown in Supplementary Figure 1. In Supplementary Figure 2, both the lumen and calcifications were segmented using our previous deep learning–based segmentation method [12]. Lumen and calcification labels were then used as the inputs of the feature extraction process.

### Supplemental results

**Supplemental Table 1.** List of the extracted features from each frame. This is a comprehensive list; recall that if you have an area, for example, we multiply it with the six statistical assessments. When we put all of this together, we get the 238 features.

| Lumen features | | Calcification features | | Statistics |
|---|---|---|---|---|
| 2D features (Frame – based) | 3D features (Lesion – based) | 2D features (Frame – based) | 3D features (Lesion – based) | |
| **Area** | **Volume** | **Max arc angle** | **Volume** | **Mean** |
| **%Area of Stenosis** | Equivalent diameter | **Max thickness** | **Volume index** | **Median** |
| **Major axis length** | Extent | **Max depth** | **Length** | **SD** |
| **Minor axis length** | Convex volume | **Area** | Equivalent diameter | Max |
| Perimeter | Solidity | **Major axis length** | Extent | Min |
| Extent | Surface area | **Minor axis length** | Convex volume | Skewness |
| Eccentricity | | Extent | Solidity | Kurtoses |
| Solidity | | Eccentricity | Surface area | |
| Circularity | | Perimeter | Number of deposits | |
| Area < 0.5*Ref | | Solidity | Calcification % | |
| Area < 0.7*Ref | | Circularity | | |
| Area < 0.9*Ref | | Stretch ration | | |

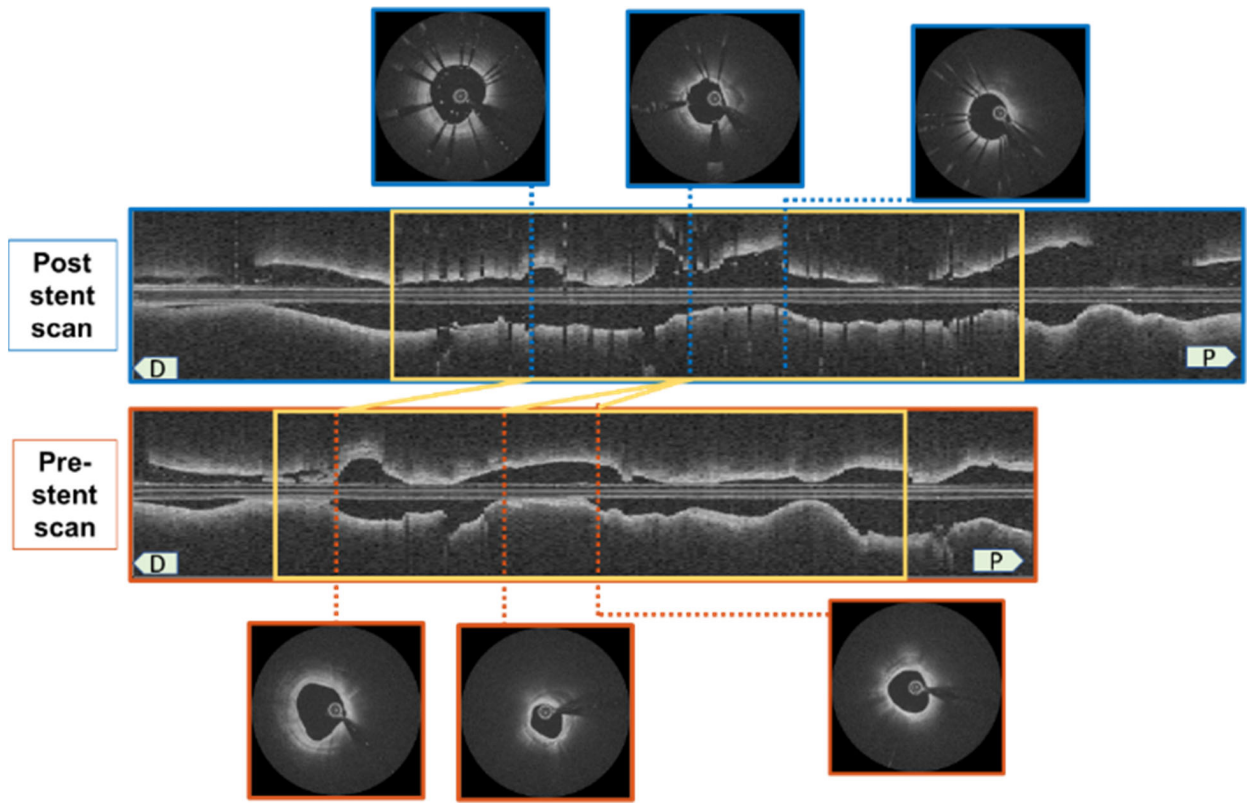

**Supplemental Figure 1.** Co-registration of pre- and poststenting IVOCT images. This is a manual process whereby landmarks (e.g., side branches and calcifications) were matched up by changing the z-offset and angle of the poststent image.

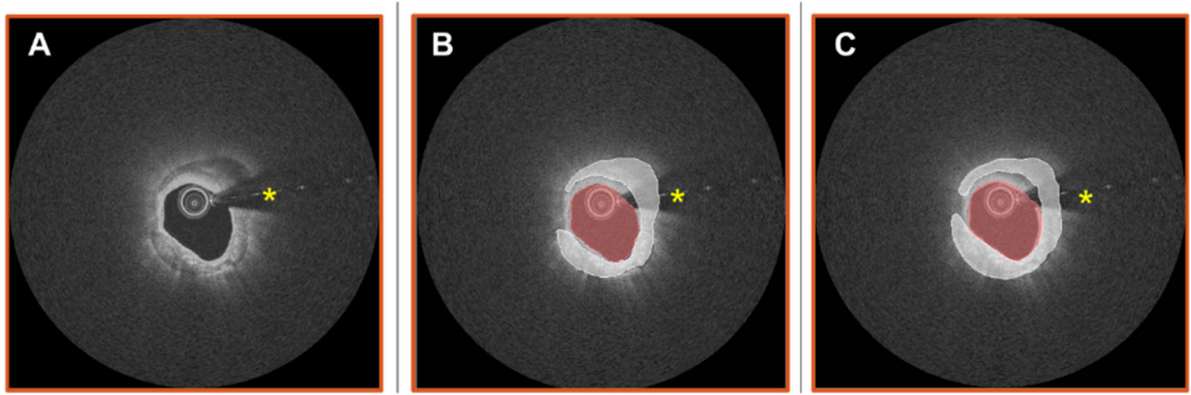

**Supplemental Figure 2.** IVOCT automated segmentation using our deep learning approach, described in (12). (A) IVOCT image. (B) Manually labeled image. (C) Automatic segmentation. The lumen is shown in red, and the calcification is shown in white. There was very good agreement between the manual and automated assessments. The asterisk (*) indicates the guidewire shadow.

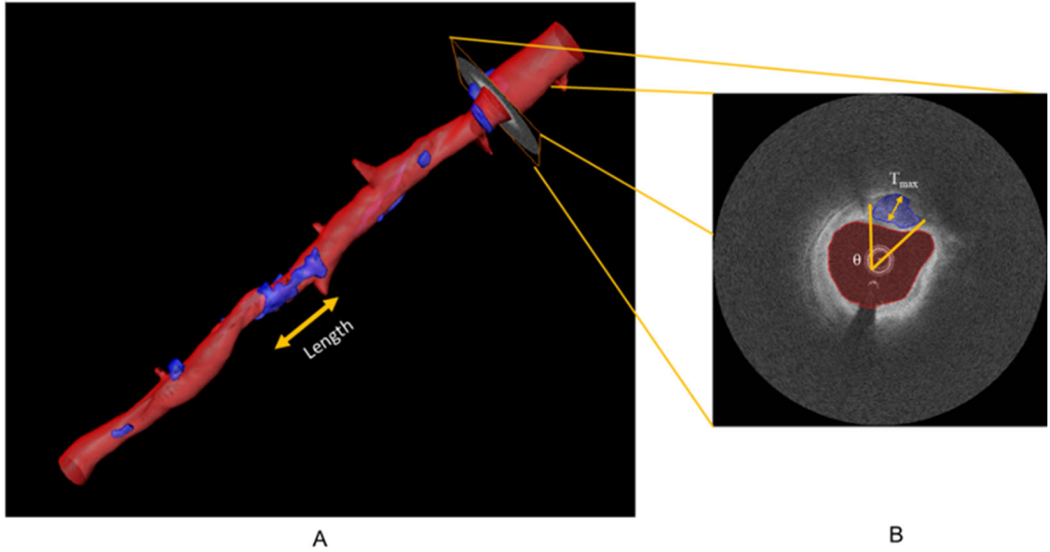

**Supplemental Figure 3.** Calcifications and their quantification. The three-dimensional rendering (A) includes multiple calcifications in blue. In the image slice (B), the calcification is tinted blue. Calcification attributes such as calcification length and calcification angle (θ) and maximum thickness ($T_{max}$) can be measured from the three-dimensional volume and the two-dimensional image frame, respectively.

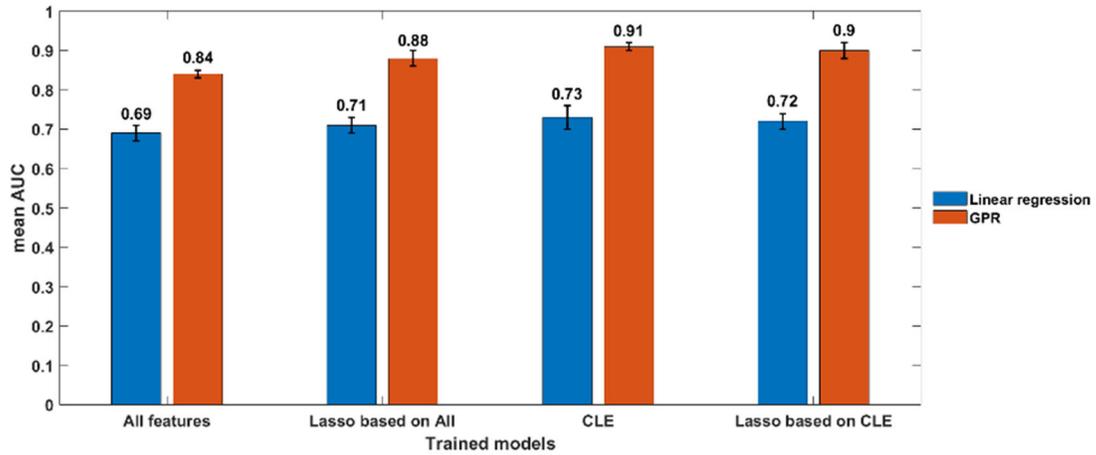

**Supplemental Figure 4.** Comparison of linear regression (LR) and Gaussian process regression (GPR) as a function of different feature groups. The mean AUC was reported as the performance metric across the fivefold validation. Features selected based on LASSO improved the performance for both LR and GPR as compared with training both models using all features. A slight degradation in performance occurred when training both models using a subgroup of the best 20 features ranked by LASSO and based on the CLE feature group. The CLE features provided the best performance for both models.

**Supplemental Table 2.** The CLE feature group ranked according to their importance as determined using LASSO on the segmental method. The calcification angle, calcification area, lumen area, percentage area stenosis (%AS) had the greatest impact on stent expansion. Calcification percentage (feature 8) is the percentage of frames with calcifications.

| # | Group | Feature | Statistic | # | Group | Feature | Statistic |
|---|---|---|---|---|---|---|---|
| 1 | Calcification | Angle | Mean | 11 | Calcification | Thick | Median |
| 2 | Lumen | Area | Mean | 12 | Calcification | Depth | Mean |
| 3 | Lumen | % AS | Median | 13 | Calcification | Thick | SD |
| 4 | Lumen | % AS | Mean | 14 | Lumen | Volume | - |
| 5 | Calcification | Area | Mean | 15 | Calcification | Depth | Median |
| 6 | Calcification | Angle | SD | 16 | Calcification | Thick | Mean |
| 7 | Lumen | % AS | SD | 17 | Calcification | Angle | Median |
| 8 | Calcification | % | - | 18 | Calcification | Area | Median |
| 9 | Calcification | Area | SD | 19 | Calcification | Volume | - |
| 10 | Lumen | Area | SD | 20 | Calcification | Depth | SD |

**Supplemental Table 3.** Summary of the performance of the analysis methods. The best performances were obtained by applying the segmental method (highlighted) with the Gaussian process regression (GPR) algorithm in combination with the CLE feature group. We also examined the effect of adding the calcification phenotype as an independent variable on the performance of the lesion-based model. We retrained the models of the lesion-based approach after adding the calcification type as an independent variable. The mean AUC was improved from 0.73 to 0.76, as indicated in the red box.

| Metric | Frame-based | Segmental analysis | Lesion-based | Lesion-based with CP |
|---|---|---|---|---|
| Accuracy | 0.78 ± 0.04 | **0.84 ± 0.04** | 0.71 ± 0.01 | 0.75 ± 0.02 |
| Sensitivity | 0.81 ± 0.06 | **0.87 ± 0.05** | 0.75 ± 0.01 | 0.75 ± 0.02 |
| Specificity | 0.75 ± 0.06 | **0.82 ± 0.05** | 0.68 ± 0.02 | 0.75 ± 0.01 |
| AUC | 0.79 ± 0.04 | **0.85 ± 0.02** | 0.73 ± 0.02 | 0.76 ± 0.02 |

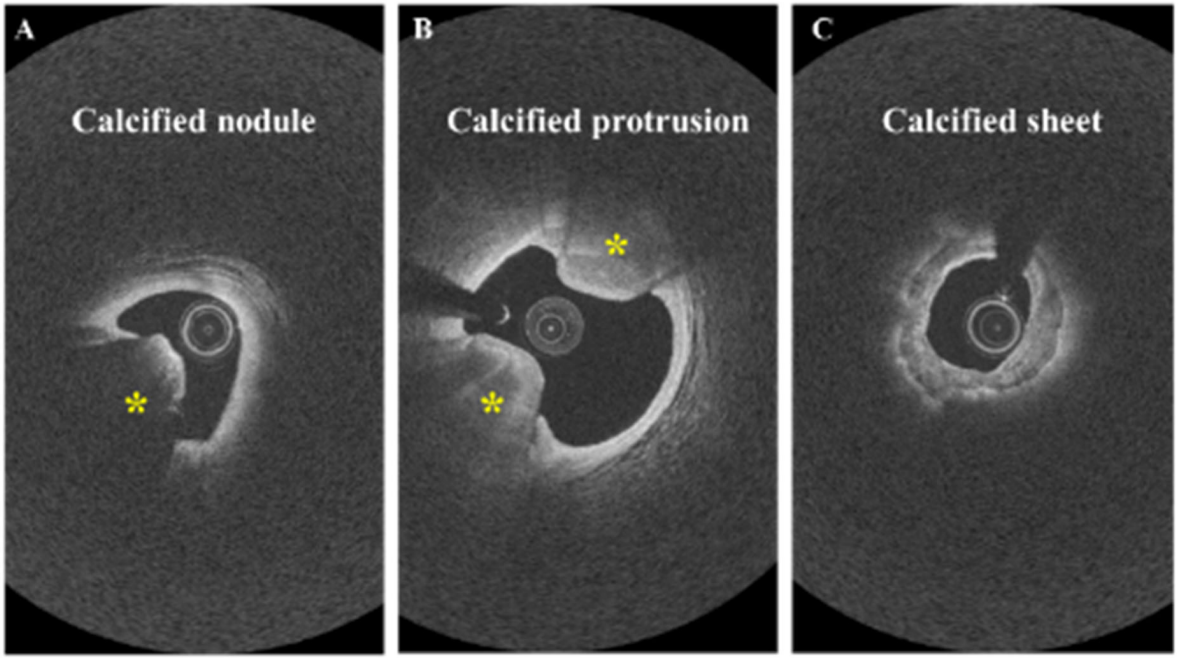

**Supplemental Figure 5.** Calcification types in the IVOCT images. (A) Calcified nodule. (B) Calcified protrusion. (C) Superficial calcific sheet.

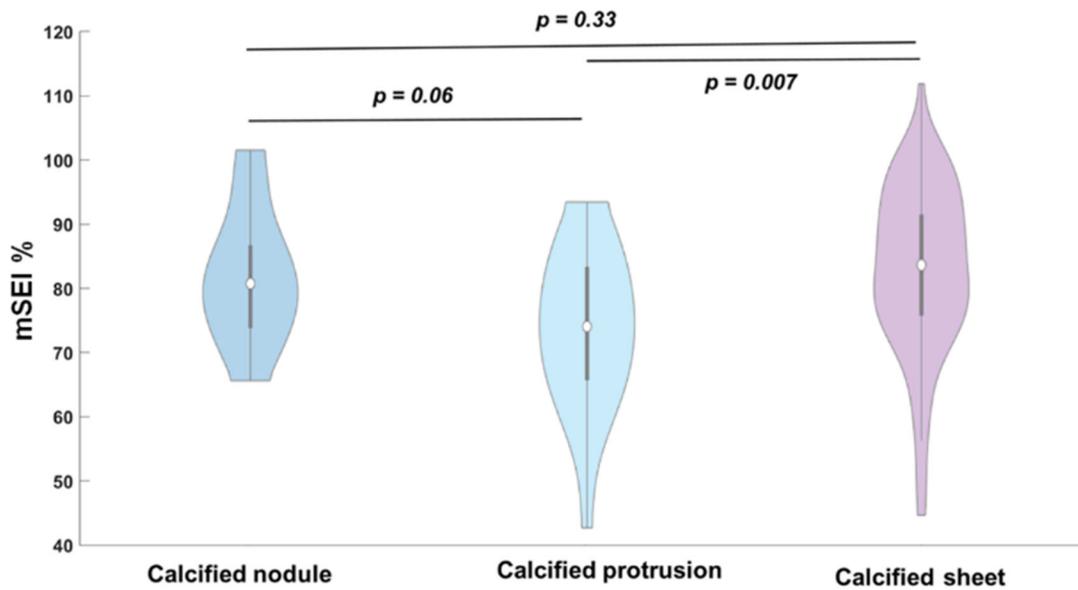

**Supplemental Figure 6.** Distribution of the SEI among different calcification phenotypes. Calcification protrusion had the lowest mSEI median value as compared with the other types.